\documentclass[preprint,preprintnumbers,amsmath,amssymb,nofootinbib]{revtex4}
\usepackage{graphicx,color}
\usepackage{amsmath,amssymb}
\usepackage{bm}
\usepackage{float}
\def\changed#1{\textcolor{black}{#1}}

\newcommand{\e}{{\mathrm{e}}}
\newcommand{\lk}{\left(}
\newcommand{\rk}{\right)}
\newcommand{\ltk}{\left\{}
\newcommand{\rtk}{\right\}}
\newcommand{\ldk}{\left[}

\newcommand{\rdk}{\right]}
\newcommand{\bk}{\bm{k}}
\newcommand{\bp}{\bm{p}}
\newcommand{\br}{\bm{r}}
\newcommand{\bq}{\bm{q}}
\newcommand{\bx}{\bm{x}}
\newcommand{\nn}{\nonumber \\}
\usepackage[dvipdfmx,
    bookmarks=true,bookmarksnumbered=true]{hyperref}

\usepackage{color}
\usepackage{ulem}

\begin{document}
\title{The ground state of polaron in an ultracold dipolar Fermi gas}
\author{Kazuya Nishimura}
\affiliation{Department of Mathematics and Physics, Kochi University, Kochi 780-8520, Japan}
\author{Eiji Nakano}
\affiliation{Department of Mathematics and Physics, Kochi University, Kochi 780-8520, Japan}
\author{Kei Iida}
\affiliation{Department of Mathematics and Physics, Kochi University, Kochi 780-8520, Japan}
\author{Hiroyuki Tajima}
\affiliation{Department of Mathematics and Physics, Kochi University, Kochi 780-8520, Japan}
\author{Takahiko Miyakawa}
\affiliation{Faculty of Education, Aichi University of Education, Kariya 448-8542, Japan}
\author{Hiroyuki Yabu}
\affiliation{Department of Physics, Ritsumeikan University, Kusatsu 525-8577, Japan}
\date{\today}

\begin{abstract}
An impurity atom immersed in {an} ultracold atomic Fermi gas {can form} a quasiparticle,
so-called Fermi polaron, due to impurity-fermion interaction.
We consider {a} three-dimensional homogeneous dipolar Fermi gas as a medium,
where the {interatomic} dipole-dipole interaction (DDI) makes the Fermi surface
deformed {into} a spheroidal shape,
and, using {a} Chevy-type variational method, investigate the ground-state
properties of the Fermi polaron: the effective mass,
the momentum distribution of {a} particle-hole ({\it p-h}) excitation,
the drag parameter,
and the medium density modification around the impurity.
These quantities are shown to exhibit spatial anisotropies {in such a
way as to reflect} the momentum anisotropy of the background dipolar Fermi gas.
We {have also given} numerical results {for}
the polaron properties at the unitarity limit
of the impurity-fermion interaction
{in the case in which} the impurity and fermion masses are equal.
{It has been} found that the transverse effective mass and the transverse momentum drag parameter of the polaron
both {tend to decrease} by $ \sim 10\%$ when the DDI strength is raised from $0$ up to around its critical value,
while the longitudinal ones exhibit a very weak dependence on the DDI.
\end{abstract}

\maketitle

\section{Introduction}
{The concept of a} polaron was originally suggested by Landau and Pekar
to describe the electron conduction in ionic crystal lattices
\cite{Zh.Eksp.Teor.Fiz16.341(1946),Zh.Eksp.Teor.Fiz.18.419(1948)}:
{When} slowly moving electrons in a lattice polarize ions around them,
the electron {dressed} in the local polarization of ions behaves as a quasiparticle,
{i.e.,} the polaron.
Nowadays, the polaron concept is used in more general cases {in which}
impurities in a degenerate medium {are accompanied by} medium {excitations
due to impurity-medium interactions.}

In {the past} decade, ultracold atoms have offered new opportunities
both in theoretical and experimental studies of {the polaron problem.  Typically,}
an impurity atom interacts with atoms in the {medium of an} ultracold Fermi/Bose gas
and forms a quasiparticle called Fermi/Bose polaron.
{In fact,} these kinds of polarons have been observed in experiments
{that utilize} radio-frequency response techniques
{(see Refs.\ \cite{PhysRevLett.102.230402,2012Natur.485..615K,Scazza_2017} for Fermi polarons
and Refs.\ \cite{2012PhRvA..85b3623C,2013PhRvL.111g0401S,2016PhRvL.117e5301H,2016PhRvL.117e5302J}
for Bose polarons).}
One of {the intriguing} features of these ultracold atomic polarons is
{experimental} controllability of the system characteristics, such as
particle statistics, interaction strength, and {dimensionality},
{which would help to verify various} theoretical studies of {the polaron problem.}
For instance, because of dominance of the low-energy scattering processes
in ultracold dilute medium, {not only are} the atomic interactions
characterized by a few scattering parameters, {namely,} the scattering
lengths and the effective ranges, {but also} their strengths are tunable
by external field \cite{RevModPhys.82.1225}.
{Another feature,} which is more characteristic {of polarons themselves,}
is the dependence {of the polaron properties on} the Fermi/Bose particle
statistics of the medium.  {In the medium of a Bose gas,} the excitations
caused by the impurity atom are Bogoliubov {phonons on top of the} Bose-Einstein condensate (BEC)
\cite{2015PhRvA..92c3612A,2013PhRvA..88e3632R,2016PhRvL.117k3002S}, {while}
in the {medium of a} degenerate Fermi gas, various kinds of particle-hole-pair
excitations {near} the Fermi surface behave as effective degrees of freedom
and interact with the impurity
\cite{PhysRevLett.97.200403,PhysRevLett.98.180402,2012PhRvA..85c3631B,Trefzger_2013}.
{Particularly in a} strong coupling regime of the impurity-medium interaction,
the impurity and the excitations {can constitute} few-body bound states,
e.g., dimer molecules, Efimov-like states, tetramers, {and so on} \cite{PhysRevA.80.053605,2015PhRvA..92c3612A,2015PhRvL.115l5302L,2016PhRvL.117k3002S,2014RPPh...77c4401M}.
%The systems of ultracold atoms can give such a unique opportunity
%for the study of a few-body problem in medium
%beyond the simple polaron picture,
%in spite that the simple quasiparticle picture should be modified drastically or even lost
%in extremely strong-coupling regime\cite{MULKERIN201929}.
%H. Tajima: I have deleted the above sentence because Ref. \cite{MULKERIN201929} shows the breakdown of the quasiparticle picture at finite temperature ($T\sim 0.8T_F$).

In the present study, {instead of treating such possible} few-body correlations
{beyond} the polaron picture,
we focus on the medium effect on quasiparticle properties of {a single} polaron
by taking {a} dipolar Fermi gas as the {surrounding} medium, {whose
salient features} are the anisotropic and long-range nature of the dipole-dipole
interaction (DDI).  In the preceding theoretical studies {based on}
a variational method {within} the Hartree-Fock (HF) approximation,
it has been shown that the DDI causes a Fermi surface deformation (FSD)
in the ground state of a spin polarized dipolar Fermi gas~\cite{PhysRevA.77.061603,Sogo_2009}.
Such an exotic state, however, may not be robust for strong DDI strengths.
{In fact,} it has been {predicted}
from {theoretical} studies of density {fluctuations in} dipolar Fermi gases
\cite{Chan_2010,Ronen_2010,Kestner_2010} that {above a critical strength,}
the FSD becomes unstable
{with respect to} growth of density waves with an infinitesimal wave vector
in the direction perpendicular to the dipoles.
{Such} instability of a dipolar Fermi gas has also been reported
{from a thermodynamic point of view based on}
the energy functional approach \cite{Sogo_2009,PhysRevA.84.053603,PhysRevA.91.023612}.
In experiments, ultracold dipolar Fermi gases have been realized
for highly magnetic atoms {and} for polar molecules having large magnetic or electric dipole moments
\cite{Lu_2012,PhysRevLett.112.010404,Naylor_2015,DeMarco_2019}, and {remarkably}
Aikawa et al.\ \cite{aikawa2014observation} has observed the FSD
in a degenerate dipolar Fermi gas of Er atoms.
%
%There are several theoretical approaches to study polaron problems, including
%perturbation theory \cite{MULKERIN201929,Trefzger_2013},
%variational theory \cite{PhysRevA.74.063628,PhysRevA.96.033627},
%Monte-Calro (MC) simulations \cite{PhysRevLett.97.200403,PhysRevB.87.115133},
%and functional renormalization group \cite{PhysRevA.95.013612}.
%Chevy considered a polaron in no interacting fermions using variational method,
%and suggested very powerful ansatz~\cite{PhysRevA.74.063628}. Chevy assumed that only one particle-hole pair are made by the impurity-fermion interaction. Chevy ansatz is simple but reproduced MC result and give correct polaron energy~\cite{2001.10450}.
%If you want to know more detail about fermi polaron,
%see a review \cite{Massignan_2014}.
%Reflecting high tunablity of ultracold atoms, many fermi polaron problems were studied
%\cite{PhysRevA.96.033627,MULKERIN201929,Trefzger_2013,PhysRevLett.102.230402}.
%However, there is no study of polaron in dipolar fermions despite of interesting properties of dipolar fermions.

Inspired by {such} theoretical and experimental progress,
%we take the dipolar Fermi gas and study the quasiparticle properties of
%{a} non-dipolar impurity atom therein
in this paper we consider a single {nondipolar} atomic impurity
moving in {a} three-dimensional homogeneous dipolar Fermi gas
at zero temperature and {then examine how its} quasiparticle
properties {such as} the effective mass {and}
the particle-hole excitations around the impurity
{depend on the direction and strength of the DDI
via the resulting FSD.}
% which exhibit spatial anisotropies
By observing the medium density distribution around the impurity,
{furthermore, we briefly} discuss a possible instability of the system
toward a density collapse, which can be triggered by impurities
through the {attraction with the dipolar gas that is already
susceptible to density fluctuations in the presence of the DDI.}
In the present calculations,
we take the unitarity limit in the impurity-medium interaction
{to} compare our results for the {spatially anisotropic medium}
with {the existing numerical data obtained} for the spatially
isotropic {medium} at the unitarity limit\cite{PhysRevA.74.063628,PhysRevLett.103.170402}.
It should be noted that
the polaron {problem} in {a} dipolar Bose gas has been studied in Refs.\
\cite{PhysRevA.89.023612,2019JPhB...52a5004A}, in which effects of the Bogoliubov
{phonons} in a dipolar BEC on the polaron dispersion relation
and spectrum intensity {have been clarified.}

In Sec.\ \ref{sec2}, we introduce the formalism {to describe}
a single atomic impurity in {a} dipolar Fermi gas
and {then} give an effective Hamiltonian of the system
{by using} the single particle energy of the dipolar Fermi gas
{obtained from} the HF calculation.
To obtain the ground state wave function,
we employ the variational method {developed}
by Chevy \cite{PhysRevA.74.063628} {in such a way as to include}
a single pair of particle-hole ({\itshape p-h}) excitations
{near} the deformed Fermi surface.
In Sec.\ \ref{sec3}, we show the numerical {results for the} polaron energy,
{the} transverse and longitudinal effective masses, the drag parameter
{that characterizes the} momentum distribution of {\it p-h} excitations,
and the medium density modification, {together with} discussions on these results.
{Section} \ref{sec4} is devoted to summary and outlook.
In this paper, we {use units in which} $\hbar = 1$.

\section{Formalism} \label{sec2}

In this section, we start with the full Hamiltonian for the system of a single impurity
in {a} uniform dipolar Fermi gas of single atomic species,
and {implement} the Lee-low-Pines (LLP) transformation \cite{Phys.Rev.90.297}
to obtain the Hamiltonian {written} in terms only of the dipolar fermions.
We then {apply} the HF approximation to the dipolar Fermi gas,
and obtain an effective Hamiltonian, {from} which we calculate various
properties of the {Fermi} polaron.

\subsection{Full Hamiltonian}

First we consider a homogeneous gas of single-component dipolar fermions at zero temperature,
whose Hamiltonian is given by
\begin{align}
    H_{d} &=\int_{\br} \psi^{\dagger}(\br) \frac{-\nabla^2}{2m} \psi(\br)
           +\frac{1}{2} \int_{r,r'} \psi^{\dagger}(\br') \psi^{\dagger}(\br) V_{dd}(\br-\br') \psi(\br) \psi(\br'),
\end{align}
where the field operator $\psi(\br)$ is for a fermion with the mass $m$, {and}
$\int_{r}$ is the abbreviation of the integration over volume $V$, $\int_V d^{3}r$.
With the strength of the dipole moment $d=|{\bm d}|$,
the DDI $V_{dd}(\br)$ is given by
\begin{align}
    V_{dd}(\br) &= \frac{d^2}{r^3} \lk 1 -3\frac{z^2}{r^2} \rk,
\end{align}
where the dipole moments are assumed to be polarized along the z-axis.
The Fourier transform of $V_{dd}(\br)$ becomes
\begin{align}
    \tilde{V}_{dd}(\bq) =\int_r V_{dd}(\br) e^{-i{\bm q} \cdot {\bm r}}
                      =\frac{4\pi}{3} d^2 \lk 3\cos^2\theta_{\bq} -1 \rk,
\end{align}
where $\theta_{\bq}$ is the angle between {the momentum transfer} ${\bm q}$
and the dipole moment ${\bm d}$.
Note that in this Hamiltonian, there is no contact interaction
between {two fermions in} the polarized (single component) medium.
{Even in the presence of such contact interaction, the Pauli exclusion
principle would allow us to ignore the resultant low-energy scattering.}

Now let us introduce a single non-dipolar impurity of the mass $M$,
which interacts with the dipolar medium fermions with a coupling constant $g$.
Then the full Hamiltonian of the impurity and the medium dipolar Fermi gas is given by
\begin{align}
    H(\bx) &=H_d
          -\frac{\nabla_x^2}{2M}
          +g \int_r \psi^{\dagger}(\br) \psi(\br) \delta(\br-\bx)
\nn
    &=\sum_{\bq} \frac{q^2}{2m} a_{\bq}^{\dagger} a_{\bq}
     +\frac{1}{2} \sum_{\bq,\bq',\bm{s}} a_{\bq'-\bm{s}}^{\dagger} a_{\bq+\bm{s}}^{\dagger} a_{\bq} a_{\bq'} \tilde{V}_{dd}(\bm{s})
\nn
    &\qquad -\frac{\nabla_x^2}{2M}
            +g \sum_{\bq,\bq'} a_{\bq}^{\dagger} a_{\bq'} e^{-i(\bq-\bq')\cdot \bx}.
\end{align}
{Here} we have used the first quantization
for the impurity in the coordinate representation {$\bx$,}
and expanded the field operator of fermions
as $\psi(\br) =\frac{1}{\sqrt{V}} \sum_{\bq }e^{i \bq \cdot \br} a_{\bq}$,
where the creation and annihilation operators $a_{\bq}^{\dagger}$ and $a_{\bq}$
satisfy the canonical commutation relation:
\begin{equation}
     \ltk a_{\bq}, a_{\bq'}^{\dagger} \rtk =\delta_{\bq,\bq'}.
\end{equation}
It should be noted that
the coupling constant $g$ is related to the $s$-wave scattering length $a$
via the Lippmann-Schwinger equation
at the low energy limit:
\begin{align}
    g^{-1} &=\frac{m_r}{2\pi  a}
            -\sum_{\bq} \frac{2m_r}{q^2},
\label{LS1}
\end{align}
where the reduced mass $m_r$ is defined
in $m_r^{-1} =m^{-1} +M^{-1}$.

\subsection{LLP transformation}

Now we apply the Lee-Low-Pines theory\cite{Phys.Rev.90.297}
to the Hamiltonian defined in the previous subsection.
We define the operator $S(\bx)$ with the momentum operator of medium fermions $\hat{\bm{P}}$:
\begin{equation}
    S(\bx) =e^{i \bx \cdot \hat{\bm{P}}} \qquad \mbox{where} \qquad
    \hat{\bm{P}} =\sum_{\bq} \bq a_{\bq}^{\dagger} a_{\bq}.
\end{equation}
{This operator generates} the LLP transformation for the annihilation operator
$a_{\bq}$ {as}
\begin{equation}
    S(\bx) a_{\bq} S^{-1}(\bx) = a_{\bq} e^{-i \bq\cdot \bx}.
\end{equation}
It should be noted that it transforms fermions to the co-moving frame of the impurity
{in which} the impurity keeps staying at the origin.
The Hamiltonian is {then} transformed as
\begin{align}
    S(\bx) H(\bx) S(\bx)^{-1}
    &=\sum_{\bq} \frac{q^2}{2m} a_{\bq}^{\dagger} a_{\bq}
     +\frac{1}{2} \sum_{\bq,\bq',\bm{s}} a_{\bq'-\bm{s}}^{\dagger} a_{\bq+\bm{s}}^{\dagger} a_{\bq} a_{\bq'} \tilde{V}_{dd}(\bm{s})
\nn
    &\qquad +\frac{\lk -i \bm{\nabla}_x -\hat{\bm{P}} \rk^2}{2M}
            +g \sum_{\bq,\bq'} a_{\bq}^{\dagger} a_{\bq'},
\label{shs}
\end{align}
{which in turn} satisfies the commutation relation:
\begin{equation}
     \ldk S(\bx) H(\bx) S(\bx)^{-1} , -i\bm{\nabla}_x \rdk=0.
\end{equation}
{This relation} implies that
the derivative operator $-i\bm{\nabla}_x$ can be replaced with a $c$-number momentum,
i.e., $-i\bm{\nabla}_x=\bm{P}$.  {This impurity momentum $\bm{P}$ corresponds to
the total momentum of the transformed system (or a polaron)
via $S(-i\bm{\nabla}_x +\hat{\bm{P}})S^{-1}=-i\bm{\nabla}_x$.}
\changed{
It should be noted that
$S H S^{-1}$ and $H$ are of unitary equivalence by the $S$ transformation,
and hence the dependence of the impurity coordinate $\bx$ in the original wave function
can always be recovered by the inverse transformation.
}

\subsection{Mean-field description of dipolar Fermi gas} \label{meanfield}

It has been shown
that a spin polarized dipolar Fermi gas takes on a deformed Fermi surface
due to the exchange contribution of the DDI~\cite{PhysRevA.77.061603}.
In the mean-field approximation,
the single-particle energy $\varepsilon_{\bq}$ of fermions {can be} obtained
from a self-consistent HF equation {as}
\begin{align}
\label{HFspe}
    \varepsilon_{\bq} =\frac{q^2}{2m}
                  -\sum_{\bq'} \tilde{V}_{dd}(\bq-\bq') n_{\bq'},
\end{align}
where the second term is the exchange contribution
with the Fermi-Dirac distribution function
given by $n_{\bq}=\theta(\epsilon_F-\varepsilon_{\bq})$
at zero temperature;
$\epsilon_F$ is the Fermi energy.
The direct contribution vanishes due to the symmetry in the momentum space.
In Ref.~\cite{Ronen_2010},
it is shown
that the Fermi surface determined {from numerical calculations} of the self-consistent single-particle energy
{can be} well described by a spheroidal form:
\begin{align}     \label{nsph}
    n_{\bq} =\theta\left( k_F^2
                     -\frac{1}{\beta} [q_x^2 +q_y^2]
                     -\beta^2 q_z^2 \right),
\end{align}
where $k_{F}$ is the Fermi momentum for non-interacting fermions.
The deformation parameter $\beta$ is in general less than one ($\beta<1$)
due to {the attractive nature} of the exchange energy,
leading to a prolate shape of the Fermi surface.
For weak DDI strengths, {$\beta$ has been calculated
in Ref.\ \cite{Ronen_2010} as}
\begin{equation}
     \beta =1 -\frac{2md^2k_F}{9\pi}.
\end{equation}

\begin{figure}[H]
\begin{center}
        \includegraphics[width=10.cm]{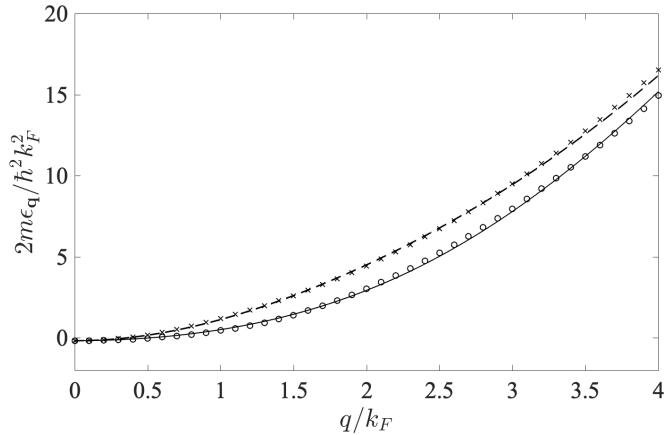}
\end{center}
\caption{Model single-particle energies $\epsilon(0,0,q)$ (solid line)
and $\epsilon(q,0,0)$ (dashed line)
{as plotted in the presence of the DDI ($C_{dd}=1.0$)} as a function of $q$.
In this calculation,
the parameters are fixed as $\beta=0.7769$, $\lambda^{2}=1.0702$, {and $k_c=2.5k_F$.}
The corresponding single-particle energies are also derived
from numerical calculations
in the HF approximation (circles and crosses).}
\label{SPE1}
\end{figure}

In this study,
instead of solving the self-consistent HF equation~(\ref{HFspe}) numerically,
we use a model single-particle energy that reproduces the deformed Fermi surface:
\begin{align}
    \label{VMspe}
    \varepsilon_{\bq} =\varepsilon_0
                  +\frac{1}{2m_t(q)}(q_x^2+q_y^2)
                  +\frac{1}{2m_z(q)} q_z^2,
\end{align}
where $\varepsilon_0 $ is an energy shift.
The momentum-dependent effective masses, $m_z(q)$ and $m_t(q)$, are given by
\begin{align}
\label{massz}
     \frac{m_z(q)}{m} &=\left( \frac{1}{\lambda^2 \beta^2} -1 \right)
                        e^{-\frac{q^2}{k_c^2}}
                       +1, \\
\label{masst}
     \frac{m_t(q)}{m} &=\left( \frac{\beta}{\lambda^2} -1 \right)
                        e^{-\frac{q^2}{k_c^2}}
                       +1,
\end{align}
where $q=|\bm q|$, {and $\lambda$ denotes}
the curvature of the single particle energy.
The parameters $\beta$, $\lambda$, and $k_c$ are determined
{from detailed calculations of the ground state properties
based on such a} variational method {as} given in Refs.~\cite{Sogo_2009,Yamaguchi_2010};
numerical values of these parameters depend on a dimensionless DDI coupling constant defined as
\begin{equation}
     C_{dd} = md^2n_f^{1/3},
\label{cdd1}
\end{equation}
where
\begin{equation}
     n_{f} = \frac{k_{F}^{3}}{6 \pi^{2}} \label{nf}
\end{equation}
is the density of {the} dipolar Fermi gas.
See TABLE~\ref{beta} for the values of $\beta$ and $\lambda^{2}$ at various $C_{dd}$'s,
where the value of $k_c$ has been determined to be $2.5k_{F}$ independently of the value of $C_{dd}$ in such a way that the model single particle energy (\ref{VMspe}) is consistent with the HF calculations.
\begin{table}[ht]
        \scalebox{1}{
        \begin{tabular}{lccccccc} \hline \hline
          $C_{dd}$ & $0$ & $0.25$ & $0.5$ & $0.75$ & $1.0$ & $1.25$ & $1.5$ \\ \hline
          $\beta$ & $1.0000$ & $0.9340$ & $0.8747$ & $0.8224$ & $0.7769$ & $0.7374$ & $0.7032$ \\ \hline
          $\lambda^{2}$ & $1.0000$ & $1.0048$ & $1.0188$ & $1.0411$ & $1.0702$ & $1.1046$ & $1.1429$ \\ \hline \hline
           \end{tabular}
        }
\caption{The values of $\beta$ and $\lambda^{2}$}
\label{beta}
\end{table}
The model single-particle energy~(\ref{VMspe}), together
with the momentum-dependent effective masses {(\ref{massz}) and (\ref{masst}),}
recovers the free-particle spectrum for large $q$; {such asymptotic behavior}
is guaranteed in the self-consistent single particle energy (\ref{HFspe}).
Fig. \ref{SPE1} {depicts} the behaviors of the model single-particle energies
for $C_{dd}=1.0$,
together with numerical results from the self-consistent HF equation;
they are found to be almost on the top of each other.
Details of how we obtain the model single-particle energy and momentum-dependent effective masses
on the basis of the variational method are given in Appendix \ref{model single particle energy}.
%
%    \begin{figure}[H]
%        \begin{center}
%        \includegraphics[width=10.cm]{SPE7.eps}
%        \end{center}
%        \caption{The same with Fig.~(\ref{SPE5}) but $k_c=2\sqrt{2}k_F$ for $C_{dd}=1.5$.}
%        \label{SPE7}
%    \end{figure}

\subsection{Effective Hamiltonian}

{Introduction} of the impurity in the medium Fermi gas brings about
the particle-hole ({\itshape p-h}) {excitations near} the deformed Fermi surface.
Now we assume that the {medium modification} can be calculated in {perturbation theory,
while basic} properties of the medium dipolar Fermi gas {are essentially unchanged.}
Then we can {construct} an effective Hamiltonian from Eq.\ (\ref{shs})
in terms only of medium fermions {as}
\begin{align}
    S(\bx) H(\bx) S(\bx)^{-1} \quad \rightarrow \quad
    H_{\mathrm eff}(\bm{P}) =\sum_{\bq} \varepsilon_{\bq} a_{\bq}^{\dagger} a_{\bq}
                       +\frac{(\bm{P}-\hat{\bm{P}})^2}{2M}
                       +g \sum_{\bq,\bq'} a_{\bq}^{\dagger} a_{\bq'},
\label{heff}
\end{align}
where we have replaced the impurity's momentum operator $ -i\bm{\nabla}_x$
by the total momentum of the system $\bm{P}$.
In the above effective Hamiltonian,
no dynamical degrees of freedom of the impurity exist \cite{PhysRevA.96.033627}.
We {can thus find the ground state of the system
from} the dipolar fermion Hamiltonian {(\ref{heff}),}
which includes a deformed single-particle energy term (the first term)
and the self-interaction {terms} (the second and the third terms).

\section{Polaron properties} \label{sec3}

In this section, using a variational method,
we obtain the ground-state wave function for the polaron with the momentum $\bm{P}$,
with which numerical results are shown for the rest energy (binding energy),
the effective masses, the {momentum} drag parameters,
and {density fluctuations} around the impurity.
Discussions are {finally} given on these results.

\subsection{Single particle-hole pair approximation}

{By following a useful approach to} the spin imbalanced Fermi gas problem \cite{PhysRevA.74.063628,PhysRevLett.98.180402},
let us {set up} a variational state $| \Psi \rangle$
{that includes excitations of} a single {\itshape p-h} pair {near} the deformed Fermi surface:
\begin{align}
    | \Psi \rangle =F_0 |fs\rangle
                   +\sum_{k>,p<} F_{\bk,\bp} a_{\bk}^{\dagger} a_{\bp} | fs \rangle,
\label{Heff1}
\end{align}
where $| fs \rangle$ is the Fermi degenerate state
{that has} fermions occupied up to the deformed Fermi surface,
$k>$ ($p<$) denotes the summation {over momenta} above (below) the Fermi surface,
and $F_0$ and $F_{\bk,\bp}$ are variational parameters that satisfy the normalization condition:
\begin{align}
     |F_0|^2 +\sum_{k>,p<}|F_{\bk,\bp}|^2 =1.
\label{nc1}
\end{align}
It should be noted that {in the absence of the DDI,}
the above simple variational ansatz (\ref{Heff1}) works
for the description of {polaronic} properties
even in the {strongly} attractive regime
close to the unitarity limit \cite{PhysRevB.77.020408,2014RPPh...77c4401M,2018NJPh...20g3048T}.

The expectation value of the effective Hamiltonian (\ref{Heff1}) is represented by
\begin{align}
    \langle H_{\mathrm eff} (\bm{P}) \rangle
       &=\sum_{\bq} \varepsilon_{\bq} \langle a_{\bq}^{\dagger} a_{\bq} \rangle
       +\frac{ \left\langle (\bm{P} -\hat{\bm{P}})^2 \right\rangle }{2M}
              +g\sum_{\bq,\bq'} \langle a_{\bq}^{\dagger} a_{\bq'} \rangle,
\label{expe1}
\end{align}
where $\langle O \rangle$ is the expectation value of the operator $O$
with {respect to} the state $| \Psi \rangle$:
$\langle O \rangle = \langle \Psi| O |\Psi\rangle$.
{Substitution of} the state (\ref{Heff1}) {leads to}
the explicit representation of each {term in the right side of Eq.\ (\ref{expe1})}
as
\begin{align}
    \sum_{\bq} \varepsilon_{\bq} \langle a_{\bq}^{\dagger} a_{\bq} \rangle
      &=\sum_{q<} \varepsilon_{\bq} \biggr( |F_{0}|^2
                                       +\sum_{k>,p<} |F_{\bk,\bp}|^2 \biggr)
       +\sum_{q>,p<} \varepsilon_{\bq} |F_{\bq,\bp}|^2
       -\sum_{k>,q<} \varepsilon_{\bq} |F_{\bk,\bq}|^2,
\\
    \left\langle ( \bm{P} -\hat{\bm{P}} )^2 \right\rangle
      &=P^2 |F_0|^2
       +\sum_{k>,p<} |F_{\bk,\bp}|^2 [ P^2-2 \bm{P} \cdot (\bk-\bp)+(\bk-\bp)^2 ],
\\
    \sum_{\bq,\bq'} \langle a_{\bq}^{\dagger} a_{\bq'} \rangle
      &=\sum_{q<} \lk |F_0|^2 +\sum_{k>,p<} |F_{\bk,\bp}|^2 \rk \nonumber
\\
      &+\sum_{k>,p<} ( F_0 F_{\bk,\bp}^* +F_0^* F_{\bk,\bp} )
       +\sum_{q>,q'>,p<} F_{\bq,\bp}^* F_{\bq',\bp}
       -\sum_{k>,q<,q'<} F_{\bk,\bq}^* F_{\bk,\bq'}.
\end{align}

\subsection{{\itshape p-h} momentum distribution and polaron energy }

We impose the normalization condition (\ref{nc1}) on the variational state
by introducing a Lagrange multiplier $\mu$
that turns out to be the energy of the {system} with the momentum $\bm{P}$.
Up to the contribution of the energy of the background dipolar fermions,
{$\mu$ reads}
\begin{equation}
     \mu=E(\bm{P})+\sum_{q<} \varepsilon_{\bq},
\end{equation}
where $E(\bm{P})$ is the energy dispersion relation of a polaron.
From the stationary condition $\delta \langle H_{\mathrm eff} -\mu \rangle=0$,
we obtain a set of eigenvalue equations:
\begin{gather}
    \frac{P^{2}}{2 M}F_{0} +g \sum_{q<} \lk F_0
                           +\sum_{k>} F_{\bk,\bq}\rk =E F_0,
    \label{eve1}
\\
    \Omega_{\bk,\bp;\bm{P}} F_{\bk,\bp} +g \lk F_0 +\sum_{q'>} F_{\bq',\bp}
                           -\sum_{q'<} F_{\bk,\bq'}\rk =E F_{\bk,\bp},
    \label{eve2}
\end{gather}
where
\begin{equation}
    \Omega_{\bk,\bp;\bm{P}} =\varepsilon_{\bk}
                   -\varepsilon_{\bp}
                   +\frac{P^2-2 \bm{P} \cdot (\bk-\bp)+(\bk-\bp)^2}{2M}
                   +g\sum_{q<} 1.
\end{equation}
From the Lippmann-Schwinger equation (\ref{LS1}),
the bare coupling constant $g$ is found to be {of} the same order of magnitude
{as} the inverse of the momentum cutoff for the summation term,
so we {can} safely drop terms {like}
$g\sum_{q'<} F_{\bk,\bq'}$ in Eq.~(\ref{eve2}) and
also $g\sum_{q<}$ in $\Omega_{\bk,\bp;\bm{P}}$ \cite{PhysRevA.74.063628}.

From the {eigenvalue} equations (\ref{eve1})--(\ref{eve2}),
we obtain the equation {that determines} the ground-state energy $E(\bm{P})$:
\begin{equation}
    E(\bm{P}) =\sum_{p<}
          \frac{1}{ \frac{m_r}{2\pi a} -\sum_{k>}
          \lk \frac{1}{E(\bm{P}) -\Omega_{\bk,\bp;\bm{P}}}
                            +\frac{2m_r}{k^2} \rk
         -\sum_{q<} \frac{2m_r}{q^2}}
         +\frac{P^{2}}{2M}.
\label{energy}
\end{equation}
{Once} the energy $E(\bm{P})$ {is determined,}
the corresponding wave functions in the momentum representation
for the {\itshape p-h}-excitation part,
the coefficients $F_{\bk,\bp}$ in (\ref{Heff1}),
are given by
\begin{align}
    F_{\bk,\bp} &%= \frac{g \chi_p}{E(\bm{P})-\Omega_{\bk,\bp;\bm{P}}}
     =F_0 \frac{g}{E-\Omega_{\bk,\bp;\bm{P}}}
          \ldk 1 -\sum_{k>} \frac{g}{E(\bm{P}) -\Omega_{\bk,\bp;\bm{P}}} \rdk^{-1}
\nn
    &=F_0
      \frac{1}{E(\bm{P}) -\Omega_{\bk,\bp;\bm{P}}} \;
      \frac{1}{\frac{m_r}{2\pi a} -\sum_{k>}
      \lk \frac{1}{E(\bm{P}) -\Omega_{\bk,\bp;\bm{P}}} +\frac{2m_r}{k^2} \rk
     -\sum_{q<} \frac{2m_r}{q^2}}.
\label{fkp}
\end{align}
The derivation of the above formula is given in Appendix \ref{Eigenvalue problem for polaron energy}.
{Physically,} $F_{\bk,\bp}$ {represents} the probability amplitude to {create a} particle
with momentum $\bk$ and {a} hole with {momentum} $\bp$ {simultaneously.}

\begin{figure}[H]
    \begin{center}
    \includegraphics[width=10.cm]{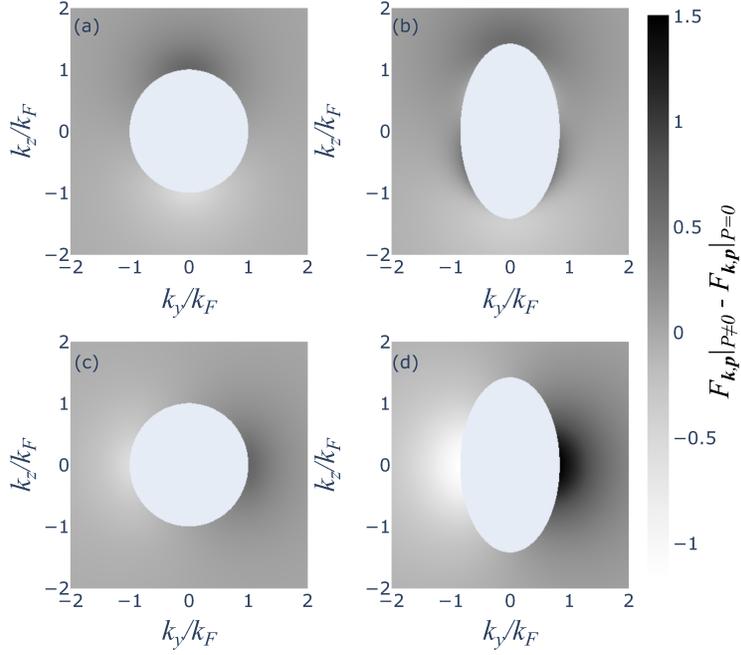}
    \end{center}
    \caption{Numerical results {for}
    	$\left.F_{\bk,\bp}\right|_{P\neq 0}- \left.F_{\bk,\bp}\right|_{P=0}$ in $k_z$-$k_{y}$ plane,
    	where $\bp \parallel \bm{P}$ and $F_{\bk,\bp}$ is averaged over $\bp$.
    	(a) $C_{dd}=0, P_z= 0.1 k_F$,  (b) $C_{dd}=1.5, P_z= 0.1 k_F$, (c) $C_{dd}=0,P_y= 0.1 k_F$,
    	and  (d) $C_{dd}=1.5,P_y= 0.1 k_F$.}
    \label{Fkp}
\end{figure}

In Fig.~\ref{Fkp}, we show numerical results {for} $F_{\bk,\bp}$ in density plots to compare
\changed{
the cases of the spherical and spheroidal Fermi surfaces,
}
i.e., the absence and presence of the DDI, for a finite $\bm{P}$.
We observe from the figures that the {\it p-h} excitations shift
\changed{in general} to the direction of $\bm{P}$
in a manner that is dependent on the deformation of the Fermi surface
and
\changed{
that,
comparing the contrasts in the density plots of panels~(b) and (d),
the {\it p-h} excitations seem easier to occur
when $\bm{P}$ is parallel to $k_y$ axis (the panel~(d)) than $k_z$ (the panel~(b))  for the finite DDI.
However, it does not immediately implies that the spatial anisotropy of the quasi-particle properties always
gets more prominent in the transverse direction than the longitudinal one,
because the longitudinal direction of the Fermi surface possesses the density of states larger
than the transverse one, which also affects the momentum integrals for physical observables.
These competitive effects result in small anisotropic effects in
the quasi-particle properties of the polaron for finite $\bm{P}$ and DDI's,
as can be seen explicitly in numerical results below.
}

\begin{figure}[H]
    \begin{center}
    \includegraphics[trim=0 50 0 0,width=13cm]{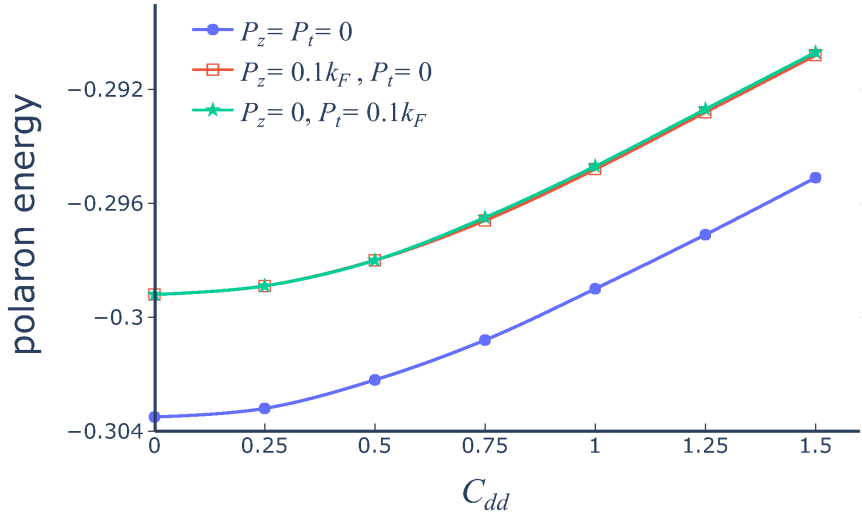}
    \end{center}
    \caption{The DDI strength $C_{dd}$ dependence of {the} polaron energy $E(\bm{P})$
    (in {units} of $k_{F}^{2}/M$) {calculated} for the total momenta
    $(P_t,P_z) =(0,0), (0,0.1), (0.1,0)$ (in {units} of $k_F$)
    at $M/m=1$ and the unitarity $a^{-1} = 0$.
    The momentum cutoff used in numerical {calculations} is $\Lambda = 100 k_{F}$,
    with which enough {convergence is} found in the momentum integrals.}
    \label{Polaronenergy}
\end{figure}

\changed{
In Fig.~\ref{Polaronenergy},
we show the numerical results for the polaron energy calculated from Eq.~(\ref{energy})
at the unitarity limit $a^{-1} = 0$ for the DDI strength $C_{dd} =0 - 1.5$,
and we also present the numerical values of the symbols in TABLE~\ref{Polaronenergy_tab}.
}

\begin{table}[ht]
        \scalebox{1.15}{
        \begin{tabular}{lccccccc} \hline \hline
            $C_{dd}$ & $ 0 $ &  $ 0.25 $ &  $ 0.5 $ &  $ 0.75 $ & $ 1.0 $ &  $ 1.25 $ & $ 1.5 $ \\ \hline
          $E(P_{t}=0,P_{z}=0)$   & $-0.3035$  & $-0.3032$  & $-0.3022$    & $-0.3008$  & $-0.2990$ & $-0.2971$ & $-0.2951$ \\ \hline
          $E(P_{t}=0,P_{z}=0.1)$ & $-0.2992$  & $-0.2989$  & $-0.2980$    & $-0.2966$  & $-0.2948$ & $-0.2928$ & $-0.2908$ \\ \hline
          $E(P_{t}=0.1,P_{z}=0)$ & $-0.2992$  & $-0.2989$  & $-0.2980$    & $-0.2965$  & $-0.2947$ & $-0.2927$ & $-0.2907$ \\ \hline \hline \end{tabular}
        }
\caption{The values of the polaron energy in Fig.\ref{Polaronenergy}.}
\label{Polaronenergy_tab}
\end{table}
We note that the case of $C_{dd} = 0$ reproduces the result
of the spin imbalanced system obtained in the literature \cite{PhysRevA.74.063628}.
\changed{
These numerical results show
that the binding energy of {the} polaron decreases with DDI strength by a small amount
and {that} the polaron energy increases with increasing {magnitude of the}
transverse and longitudinal momenta, $P_t$ and $P_z$, also by a small amount.
}
We {will} examine the momentum-dependence of the polaron energy
by evaluating the effective mass {in the next subsection}.

\subsection{EFFECTIVE MASS}

The effective mass,
which characterizes the mobility of the polaron in the medium,
has directional dependence because of the deformed shape of the Fermi surface
by {the} DDI.
In this paper, we {adopt} the definition of the effective mass
{based on} the momentum expansion of the polaron energy:
\begin{align}
      E(\bm{P}) = E(0)
           +\frac{P_{t}^{2}}{ 2 M_{t}}
           +\frac{P_{z}^{2}}{ 2 M_{z}}
           +{\mathcal O}(P^4),
\label{effective energy}
\end{align}
%
%which depends we write the energy of polaron using effective mass $M_{t},M_{z}$
%
%  On the other hand,  Taylor expansion of the polaron energy for small $P$ is
%
% \begin{align}
% E(\bm{P}) \approx E(0)
%           + \frac{1}{2} \left. \frac{\partial^{2} E}{\partial P_{t}^{2} }\right|_{P=0} P_{t}^{2}
%         + \frac{1}{2}\left. \frac{\partial^{2} E}{\partial P_{z}^{2} }\right|_{P=0} P_{z}^{2}.\label{T expansion energy}
%\end{align}
%
where $M_t$ and $M_z$ are the transverse and longitudinal effective masses,
respectively:
%  To compare eq.(\ref{effective energy}) and eq.(\ref{T expansion energy}), we obtain effective mass of polaron:
%
\begin{gather}
      M_{t}^{-1} =\left. \frac{\partial^{2} E}{\partial P_{t}^{2}} \right|_{P=0},
%
% =
% \frac{
%       \frac{-(1+R)}{4 \pi M} \int \dc p   \gbl \frac{2}{G^{3}} \gl \frac{\partial G }{\partial P_{t}} \biggr)^{2} - G_{t}^{(2)} \gbr \biggr|_{P=0}
%     }{
%       1 + \frac{(1+R)^{2}}{2 \pi }   \int \dc p \int d^{2} k \frac{F}{G^{2} H^{3/2} }    \biggr|_{P=0}
%      }
%   +  \frac{1}{M}
%
\\
%
%\mbox{and } \    &
%
     M_{z}^{-1} =\left. \frac{\partial^{2} E}{\partial P_{z}^{2}}  \right|_{P=0}.
%
% =
% \frac{
%       \frac{-(1+R)}{4 \pi M} \int \dc p   \gbl \frac{2}{G^{3}} \gl \frac{\partial G }{\partial P_{z}} \biggr)^{2} - G_{z}^{(2)} \gbr \biggr|_{P=0}
%     }{
%       1 + \frac{(1+R)^{2}}{2 \pi }   \int \dc p \int d^{2} k \frac{F}{G^{2} H^{3/2} }    \biggr|_{P=0}
%      }
% +  \frac{1}{M}.
%
\end{gather}

\begin{figure}[H]
    \begin{center}
    \includegraphics[trim=0 50 0 0,width=13cm]{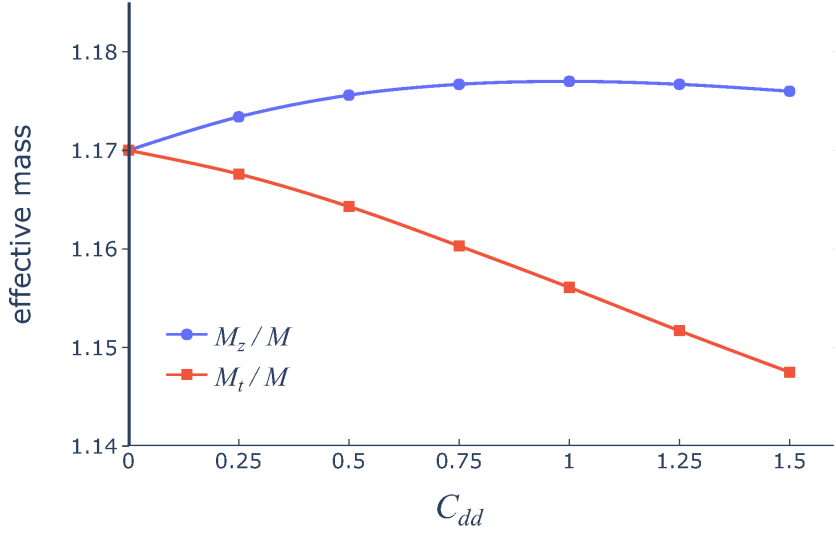}
    \end{center}
    \caption{The DDI strength $C_{dd}$ dependence of {the relative} effective masses,
    $M_z/M$ and $M_t/M$.
    The parameters {that characterize the polaronic state are set in} the
    the same {way as} in Fig.~\ref{Polaronenergy}. }
    \label{effectivemass}
\end{figure}

\changed{
In Fig.~\ref{effectivemass},
we show the numerical results
for the DDI dependence of the effective masses
and present the numerical values of the symbols in TABLE~\ref{effectivemass_tab}.
We have found from these results that
}
the longitudinal mass is not so sensitive to {the} DDI,
while the transverse one tends to decrease monotonically with the strength of the DDI.
It is not straightforward to understand such directional dependence because
whereas a polaronic cloud composed of excitations around the impurity generally acts to
increase the polaron effective mass, additional effects due to the deformation of
the Fermi surface produce anisotropy in the effective mass not only via change in the structure of
the polaronic cloud as depicted in Fig.\ \ref{Fkp}, but also via change in the
density of states available for particles and holes in the medium.
It should be noted that {such an} anisotropic effective mass has also been reported
for Bose {polarons in} a dipolar BEC \cite{PhysRevA.89.023612}.

\begin{table}[h]
      \scalebox{1.25}{
      \begin{tabular}{lccccccc} \hline \hline
        $C_{dd}$& 0 & 0.25 & 0.5& 0.75 & 1.0& 1.25 & 1.5\\ \hline
        $M_{z}/M$ & 1.170 & 1.173 & 1.176 & 1.177 & 1.177 & 1.177 &  1.176 \\ \hline
        $M_{t}/M$ & 1.170&1.168 &1.164&1.160 & 1.156 &1.152& 1.140\\ \hline
        $M_{z}/M_{t}$ & 1.000& 1.004 &1.010& 1.014 &1.018 & 1.021&1.030 \\ \hline \hline
      \end{tabular}
      }
\caption{The values of the effective masses in Fig.\ref{effectivemass} and the ratio $M_{z}/M_{t}$. }
\label{effectivemass_tab}
\end{table}
%---------------------------------------------------------
%---------------------------------------------------------
%

\subsection{Momentum drag parameter}

The change of the {polaron} effective {masses from the impurity mass $M$}
is {attributable} to the {\itshape  p-h} excitations around the impurity,
since they contribute to the kinetic term in (\ref{expe1})
through the expectation value of the momentum {of medium fermions:}
\begin{align}
    \langle \hat{\bm{P}} \rangle &=\sum_{\bq} \bq \langle a_{\bq}^{\dagger} a_{\bq} \rangle
                             =\sum_{k>,p<} (\bk-\bp) |F_{\bk,\bp}|^2.
\end{align}
{Then, let us define} the drag parameter $\eta_{ij}$ {via}
\begin{align}
      \langle \hat{\bm{P}} \rangle_i &= \eta(\bm{P})_{ij} P_j.
\end{align}
{In this way,} $\eta_{ij}$ measures
the momentum-share of the {\itshape  p-h} excitations {accompanying the polaron}
via the wave function (\ref{fkp}) that depends on $\bm{P}$ implicitly.
\footnote{
Note that, since the polaron momentum is not diffusing away
\changed{
at zero temperature in this study,
we use the term ``drag'' in the sense of the linear momentum drag by the virtual {\itshape  p-h} cloud
in the polaron,
instead of a spin drag at finite temperature \cite{2012PhRvA..85c3631B}.
It should not also be confused with the non-dissipative momentum drag/entrainment in Andreev-Bashkin effect,
which accounts for an induced super current between two different super fluids by the interaction between them
\cite{threevelocity,andreevbashkin}.
}
}
{On the basis of} symmetry considerations,
we may assume the diagonal form for the drag parameter:
\begin{equation}
    \eta_{i} P_{i} =\langle \hat{\bm{P}} \rangle_{i}
                   =\sum_{k>,p<} (k_{i} - p_{i} ) | F_{\bk,\bp}  |^{2},
\label{drag}
\end{equation}
where $\eta_{t}$ and $\eta_{z}$ are the transverse and longitudinal drag parameters.

TABLE~\ref{Drag parameter} shows the DDI dependence of the drag parameters;
\begin{table}[ht]
        \scalebox{1.25}{
        \begin{tabular}{lcccc} \hline \hline
          $C_{dd}$& $0$ & $0.5$ & $1.0$ & $1.5$ \\ \hline
          $\eta_{z}(P_{t}=0,P_{z}=0.1) $ & 0.145  & 0.149 & 0.150 & 0.150 \\ \hline
          $\eta_{t}(P_{t}=0.1,P_{z}=0) $ & 0.145  & 0.142 & 0.135 & 0.129  \\ \hline \hline
        \end{tabular}
        }
\caption{The DDI strength $C_{dd}$ dependence of the drag parameters
{calculated} for the momentum $(P_t,P_z) =(0,0.1), (0.1,0)$ {(in units of $k_F$).}
The parameters {that characterize the polaronic state are set in} the
the same {way as} in Fig.~\ref{Polaronenergy}. }
\label{Drag parameter}
\end{table}
the {\itshape p-h} {excitations contribute}
to {$13 \%$--$15 \%$ of} the momentum {of a} slowly-moving polaron.
{Just like the case} of the effective mass,
the transverse drag parameter monotonically decreases with increasing $C_{dd}$
while the longitudinal one is less sensitive {to $C_{dd}$}
than the transverse one.

\subsection{Medium density distribution}\label{sec3-3}

In the subsections above, we have studied the DDI-dependence of the polaron properties.
As suggested in {earlier investigations}, however,
the dipolar Fermi gas itself may be unstable toward a density collapse
{for sufficient strong} DDI's.
The critical value {of the} DDI strength for the collapse is estimated to be $C_{dd}\simeq 2.4$
from the compressibility {analysis} in the variational method~\cite{Sogo_2009},
{as well as} $C_{dd}\simeq 1.03$ from the analyses of collective modes~\cite{Chan_2010}
and zero sounds~\cite{Ronen_2010}.
In the present study, we cannot strictly discuss {the possibility that}
such instability {could be} caused by a impurity injection
into {a marginally} stable dipolar Fermi gas,
since fermion-density {fluctuations, which tend to develop in the
presence of the DDI, are} not incorporated in the calculation.
Instead, we evaluate {an enhancement} of the medium density distribution
around the impurity by treating the impurity as an external probe at position $\bx$
and {calculating} the medium density distribution
{as a function of} the relative distance from the impurity for $\bm{P}=0$
{as} \cite{2016PhRvB..93t5144N,PhysRevA.99.023601}
\begin{align}
    n(\br-\bx)
    &=\langle S^{-1} \psi^{\dagger}(\br) \psi(\br) S \rangle
\nonumber
\\
    &=\frac{1}{V} \sum_{\bq_1 ,\bq_2}
      \e^{-i (\bq_2 - \bq_1 ) \cdot (\br-\bx)}
      \biggr[ |F_0|^2 \langle  a^{\dagger}_{\bq_1} a_{\bq_2} \rangle
             +\lk F_0 \sum_{k>,p<}
              F_{\bk,\bp}^{*} \langle a^{\dagger}_{\bp} a_{\bk}a^{\dagger}_{\bq_1} a_{\bq_2} \rangle
             +\hbox{h.c.}\rk
\nonumber
\\
    &\qquad \qquad \qquad \qquad \qquad \qquad \qquad \quad
             +\sum_{k>,p<} \sum_{k'>,p'<}
              F_{\bk,\bp} ^{*}  F_{\bk',\bp'}
              \langle a^{\dagger}_{\bp}
                      a_{\bk} a^{\dagger}_{\bq_1}
                      a_{\bq_2} a^{\dagger}_{\bk'} a_{\bp'} \rangle
      \biggr]
\nonumber
\\
    &=n_f
     +|F_0|^2 \int_{k>,p<}
      \bar{F}_{\bk,\bp}^{*}
      \biggr[ 2 \e^{-i (\bp-\bk) \cdot (\br-\bx)}
             +\int_{k_{2}>} \bar{F}_{\bk_{2},\bp}
              e^{-i (\bk_{2}- \bk )\cdot (\br-\bx)}
\nonumber
\\
     &\qquad \qquad \qquad \qquad \qquad \qquad
      \qquad \qquad \qquad
            -\int_{p_{2}<} \bar{F}_{\bk,\bp_{2}}
              \e^{-i (\bp - \bp_{2})\cdot (\br-\bx)}
       \biggr],
\label{imferdis1}
\end{align}
where we have used the normalization condition (\ref{nc1}),
and introduced the scaled variables $\bar{F}_{\bk,\bp} =V F_{\bk,\bp}/F_0$ and
the abbreviated notation for the integrals,
$\int_{\bq} \equiv \int {\rm d}^3q/(2\pi)^3$, {etc.}
\changed{
The distribution (\ref{imferdis1}) is equivalent to the impurity-fermion density correlation function calculated,
for instance, in \cite{residue,imptrapfermion},
which gives a Friedel oscillation in the spherically symmetric case.
}
It should be noted that $F_{\bk,\bp}$ is real
and even for the transformation $k, p \rightarrow -k, -p$.
Since the density distribution $n(\br)$ diverges at $\br = 0$
in the limit of the infinite momentum cutoff for $k>$ integrals,
we have introduced a cutoff $\Lambda=2 k_{F}$ in numerical {integration,}
which is related to the effective range $r_{e} =\frac{4}{\pi \Lambda}$
in the {impurity-medium} interaction
(see Appendix \ref{effective range} for {details}).

{Fig.}~\ref{zdens} {displays} the medium density distribution
along $z$-axis and $y$-axis.
\begin{figure}
  \vspace{3cm}
  \begin{center}
    \begin{tabular}{c}
      \begin{minipage}{0.5\hsize}
        \includegraphics[bb=0 50 500 200,scale=0.45]{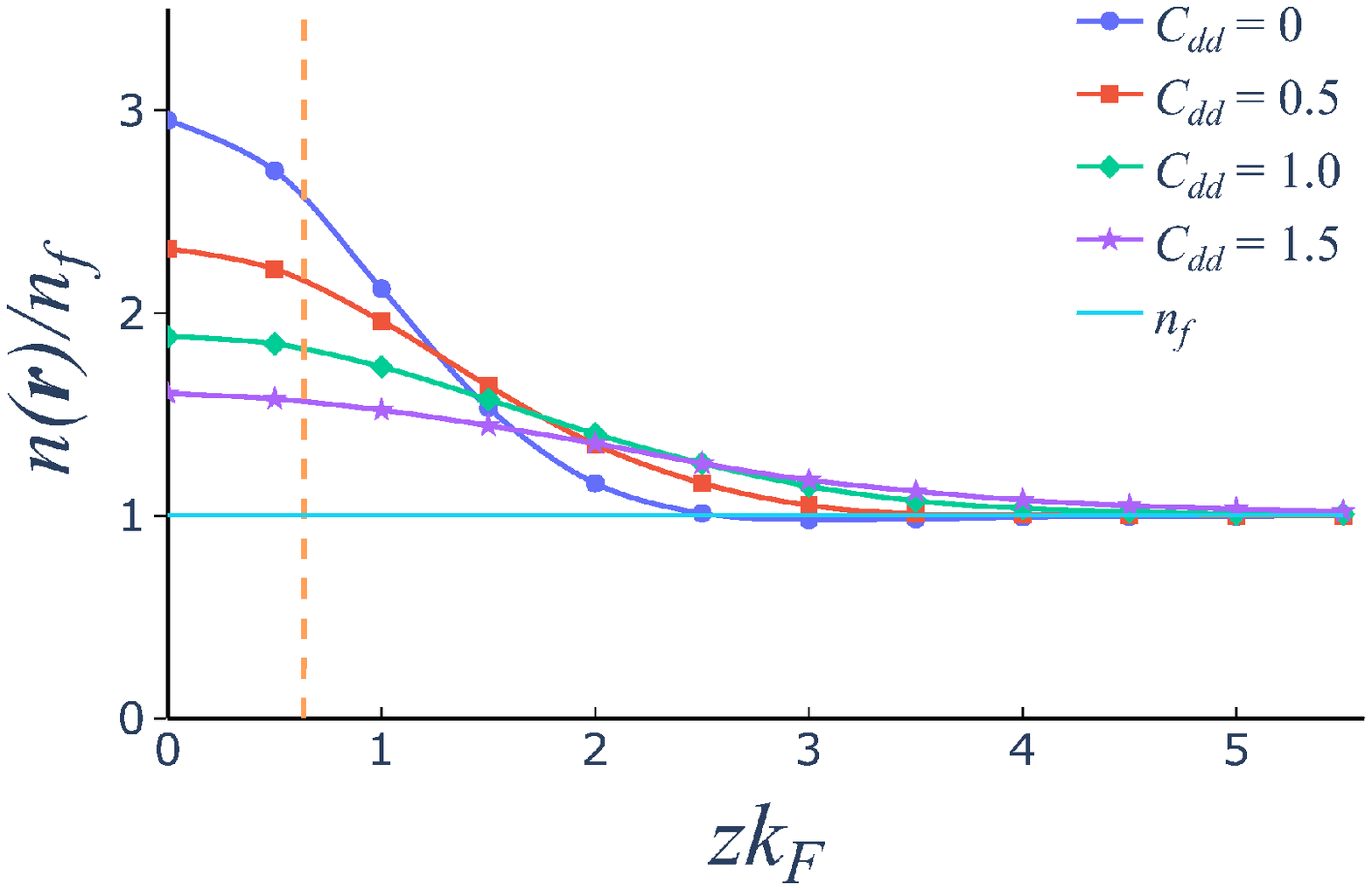}
      \end{minipage}

      \begin{minipage}{0.5\hsize}
        \includegraphics[bb=0 50 500 200,scale=0.45]{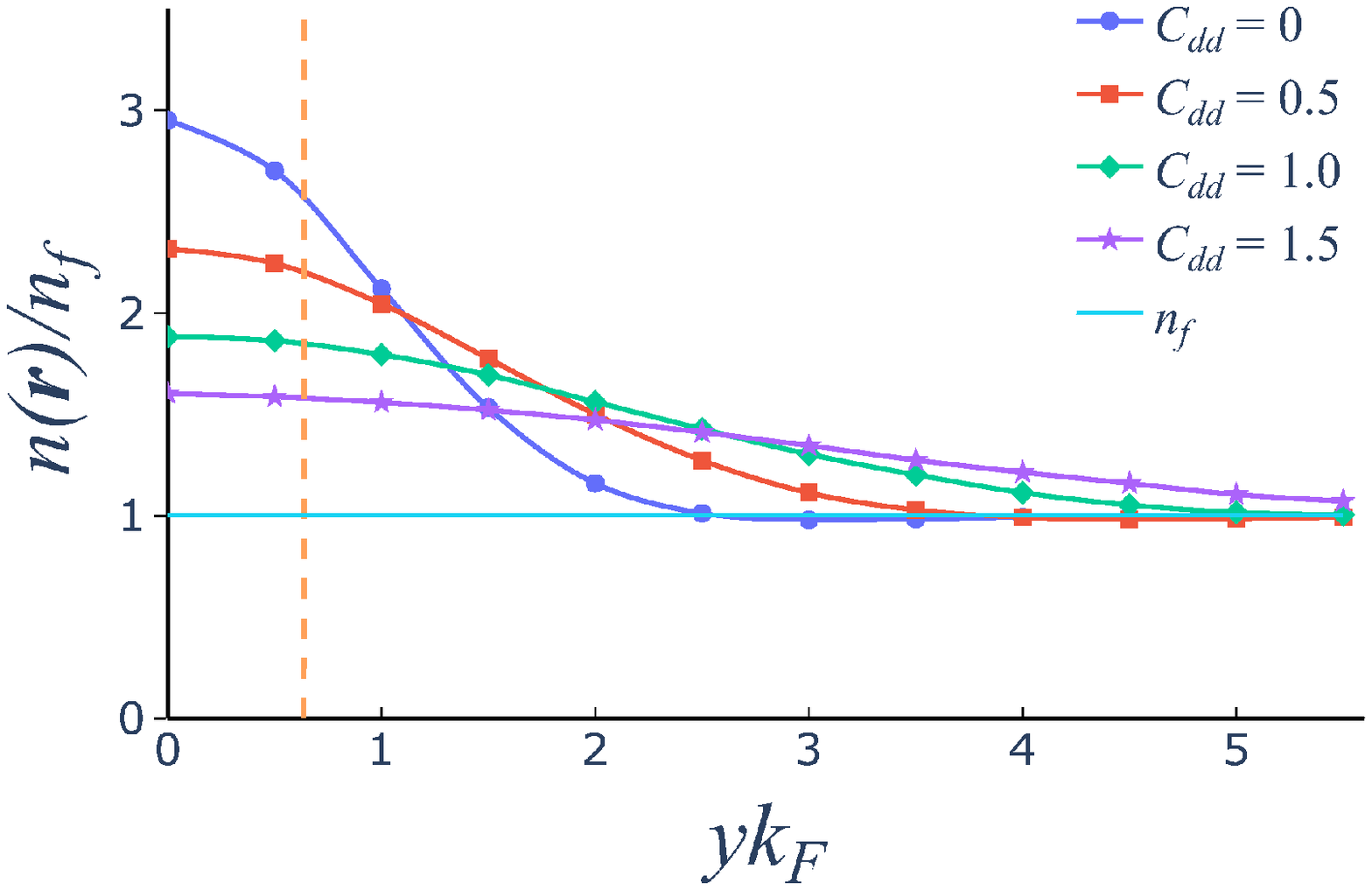}
      \end{minipage}
    \end{tabular}
  \end{center}

\caption{The medium density distribution for the longitudinal coordinate $z$ (left) and the vertical coordinate $y$ (right). {The dashed line denotes the effective range.}
The parameters {that characterize the polaronic state} are
$\Lambda = 2 k_{F}$, $M/m=1$,
$a^{-1} = 0$, and $\bm{x}=0$.
Note that $n(\br)$ approaches $n_f$ {defined in Eq.(\ref{nf})}, {which corresponds to the density of a homogeneous system in the absence of the impurity,}
when $|\br| \to \infty$. }
\label{zdens}
\end{figure}
It is found {from} these figures that
the medium density has a significant enhancement
at the center of {the Fermi} polaron due to the attractive nature of the interaction,
which is expected if the polaron is treated as a wave packet or a dilute gas {cloud}
in reality, and {that} the enhanced area broadens as the Fermi surface deformation is enlarged for {stronger} DDI's.
\footnote{
In the case of Bose polarons,
feedback {effects} of {the} impurity {on the local structure of} the medium
{have} been discussed {both} in the cases of repulsive and attractive interactions \cite{PhysRevA.99.023601,PhysRevA.100.023624}.
}
\footnote{
It should be noted that
since the average of the medium density over the impurity position $\bx$ {leads inevitably} to the uniform density,
that is,
$V^{-1}\int_x n(\br-\bx)=n_f$,
the medium density {has to undergo a Friedel-like oscillation} around $n_f$ at long distances $zk_F, yk_F \gg 2$
although its amplitude becomes very small numerically in accordance with the behavior of $F_{\bk,\bp}$.
}
{It is interesting to note that the polaronic cloud is relatively narrow in the elongated direction
of the Fermi surface, a feature that is consistent with the uncertainty principle.}

{Such a} density enhancement implies {that the
increase in} DDI strength $C_{dd}$ {acts to} expand the local area of
{the polaronic cloud and may eventually}
%so the interfusion of a cloud of impurity
trigger a local instability of {a} dipolar Fermi gas,
{which is different from} the instabilities {predicted}
in Refs.\ \cite{Chan_2010,Ronen_2010,Kestner_2010,PhysRevA.84.053603,PhysRevA.91.023612}
due to {growth of} density fluctuations in the long wavelength limit.
Unlike droplet formation in {a} dipolar BEC \cite{Wenzel_2018}, {however,}
{such a local} instability does not immediately lead to droplet formation {in} the dipolar Fermi gas.
{This is partly} because in the present model Hamiltonian there is no genuine short-range {repulsion}
among medium-fermions
%or between impurity and medium fermions,  <- g<0 is to be assumed
that can support the droplet.
%  {and partly because even in the presence of the short-range repulsion,
% the Pauli principle does not allow it to work.}

\section{Summary and outlook} \label{sec4}

We have studied the single polaron properties in a homogeneous dipolar Fermi gas
using {a} Chevy-type variational method.
The dipole-dipole interaction (DDI) in {the Fermi gas medium}
causes a spheroidal deformation of {the} Fermi surface,
which in turn leads to {anisotropy} in {the} polaron's dispersion relation.
We have investigated the DDI-dependence of {various} polaron properties
at the unitarity limit of the impurity-medium interaction;
the obtained results are summarized
{in Figs.~\ref{Polaronenergy}--\ref{zdens} and {in} TABLES~\ref{Polaronenergy_tab}--\ref{Drag parameter}}.
{We} have found significant deviations from the case of no DDI:
{The} effective mass and the momentum drag parameter for the transverse
direction {with respect to the polarization}
both decrease with increasing the strength of DDI,
while those for the longitudinal direction
are both less sensitive to the {DDI strength} than the transverse ones.
All these behaviors are attributed to {the influence of the deformation of the
Fermi surface, which is elusive because such deformation affects not only the
structure of the polaronic cloud as depicted in Fig.\ \ref{Fkp},
but also the density of states available for particles and holes in the medium.}

An interesting future direction related to the present work
is the instability problem of {a} dipolar Fermi gas,
{which could be} caused by the impurity injection.
As shown in Sec.\ \ref{sec3-3},
the impurity causes local enhancement of the medium density,
which may {eventually lead to} a density collapse. Thus,
if some repulsive forces exist to {keep} the system
{metastable against} the collapse, a droplet formation
{might be} expected \cite{PhysRevLett.89.050402}.
%The detailed study in this direction requires the calculation of a few-body correlation energy
%in dipolar Fermi gas \cite{1971RvMP...43..479B},
%including the contribution from the impurity cloud.
{Another possible} direction of study is related to the correlation among
impurities just mentioned above.
{Non-local} effective interaction between two impurities, {which can appear
via, e.g.,} a long-range {force} mediated by fluctuations
in the medium
\cite{2012PhRvA..85a3605G,2016PhRvB..93t5144N,2018JPSJ...87d3002N,2018PhRvL.121a3401C,2020PhRvL.124p3401E},
{is expected to provide} a basis
for further study of multi-impurity systems \cite{2008PhRvL.100x0406B,2011PhRvA..84e1605G,PhysRevA.102.051302,Mistakidis_2019}.
%Once the formulation for many-body description of impurity in dipolar Fermi gas is made,
%it is also interesting to study the collective nature, e.g.,
%hydrodynamic behaviors governed by the effective interaction of impurities
%\cite{2008PhRvL.100x0406B,2011PhRvA..84e1605G,PhysRevA.102.051302}.

\acknowledgments
We are grateful to Junichi Takahashi and Tomohiro Hata for useful discussion.
This work was in part supported by Grants-in-Aid for Scientific
Research through Grant Nos.\ 17K05445, 18H05406, 18H01211,
and 18K03501,
provided by JSPS.

\bibliographystyle{unsrt}
\bibliography{myrefs}

\appendix
  \renewcommand{\theequation}{A.\arabic{equation} }
  \setcounter{equation}{0}
  \section{Model single particle energy} \label{model single particle energy}
In this appendix we examine the model single particle energy (\ref{VMspe}) that approximates the self-consistent
single particle energy (\ref{HFspe}).
We {first summarize a variational method used in Sec.\ \ref{meanfield} by writing down}
the number-conserving variational ansatz for the distribution function,
\[
    n_{\bq}=\theta\left(k_F^2-\frac{1}{\beta}[q_x^2+q_y^2]-\beta^2 q_z^2 \right)
\]
with $k_{F}=(6\pi^2n_f)^{1/3}$, {which is given by Eq.\ (\ref{nsph}). Here,}
the parameter $\beta$ characterizes the deformation of the Fermi surface.
Given the ansatz, the total energy can be derived as
\begin{equation}
\label{varE}
   \frac{E(\beta)}{V}=\frac{1}{m} n_f^{5/3}\left[ C\left(\frac{2\beta}{3}+\frac{1}{3\beta^2}\right)-\frac{\pi}{3} C_{dd} I(\beta) \right]=\frac{1}{m} n_f^{5/3}{\mathcal E}(\beta),
\end{equation}
where $C=3(6\pi^2)^{2/3}/10$ and the function $I(\beta)$ is given by
\[
   I(\beta) =
   \displaystyle \frac{6}{1-\beta^3}\left[ 1-\sqrt{\frac{\beta^3}{1-\beta^3}} \arctan{\left(\frac{1-\beta^3}{\beta^3}\right)} \right]-2.
\]
{In} this variational method, the ground state {energy and the optimal value of $\beta$ are}
determined by the stationary condition for the total energy, Eq.\ (\ref{varE}),
with respect to $\beta$ {at fixed $C_{dd}$.}
For non-interacting fermions ($C_{dd}=0$), $\beta$ equals {unity}, leading to {a} spherical Fermi surface.
As the DDI strength increases, $\beta$ becomes less than one, leading to a prolate Fermi surface.

{Let us turn to the model} single particle energy {consistent with the ansatz (\ref{nsph}),
which is written in terms of the optimal $\beta$ as}
\begin{equation}  \label{spebeta}
   \epsilon_{\bm q} = \epsilon_0 +\frac{\lambda^2}{2m}  \left( \frac{1}{\beta}(q_x^2+q_y^2)+\beta^2 q_z^2 \right),
\end{equation}
{where}
\begin{equation} \label{seHF0}
   \epsilon_0 = -\frac{(6\pi^2)^{1/3}}{9\pi}\frac{k_{F}^2}{m} C_{dd} I(\beta)
\end{equation}
{is the HF self-energy at ${\bm q}=0$ obtained by substituting the ansatz (\ref{nsph}) into the right side of
Eq.\ (\ref{HFspe}).  Here,}
$\lambda$ is the curvature parameter of the single particle energy and is determined by the relationship
\[
   \epsilon_F=\epsilon_0+ \frac{\lambda^2 k_{F}^2}{2m},
\]
{where $\epsilon_F$ is given by $V^{-1} \partial E/ \partial n_f|_V$ with $E$ given by Eq.\ (\ref{varE}).}
Thus the variational single particle energy in this procedure {can be} regarded as that of a quasiparticle
with anisotropic effective masses $m_z/m=1/\beta^2\lambda^2$ {and $m_t/m = \beta/\lambda^2$ because
expression (\ref{spebeta}) can now be rewritten as}
\begin{equation}
	\label{VMspeapp}
   \epsilon_{\bm q} = \epsilon_0 + \frac{1}{2m_t}(q_x^2+q_y^2) +\frac{1}{2m_z} q_z^2,
\end{equation}
{where $\epsilon_0$ is still given by Eq.\ (\ref{seHF0}).}

\begin{figure}[H]
        \begin{center}
        \includegraphics[width=8.cm]{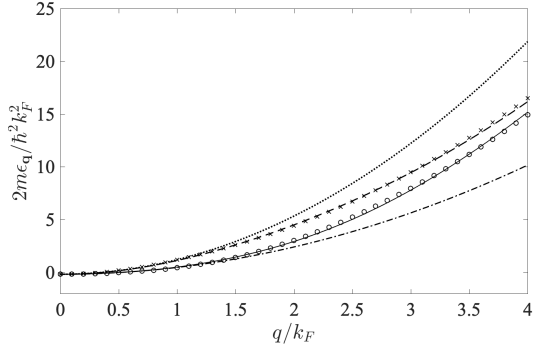}
        \end{center}
        \caption{{Same as Fig.\ \ref{SPE1} with additional plots of} $\epsilon(0,0,q)$ (dot-dashed line) and $\epsilon(q,0,0)$ (dotted line)
                 {given by} Eq.\ (\ref{VMspeapp}).}
%        \caption{Single particle energies: $\epsilon(0,0,q)$ (dot-dashed line) and $\epsilon(q,0,0)$ (dotted line) of Eq.\ (\ref{VMspeapp})
%        and $\epsilon(0,0,q)$ (solid line) and $\epsilon(q,0,0)$ (dashed line) of Eq.\ (\ref{VMspe}) with $k_c=2.5$
%        as a function of $q$ for $C_{dd}=1.0$.
%        The corresponding single particle energies are also derived from numerical calculations in the HF approximation
%        (circles and crosses).}
        \label{SPE1app}
    \end{figure}

For $C_{dd}=1.0$, the deformation and curvature parameters {can be} obtained
{from the above procedure as} $\beta=0.7769$ and $\lambda^{2}=1.0702$.
The variational single particle energies, $\epsilon(0,0,q)$ and $\epsilon(q,0,0)$,
{calculated from Eq.\ (\ref{VMspeapp}) for such parameter values}
are plotted as dot-dashed line and dotted line in Fig.~\ref{SPE1app}, respectively.
{Both of them agree} with the self-consistent HF single particle energy below and around {the} Fermi energy,
but clearly {deviate from} the HF result well above {the} Fermi energy.
{This deviation can be corrected in such a way as to reproduce} the free-particle spectrum for large $q$,
{which is guaranteed by} the self-consistent HF equation.  Thus we employ the model single particle
energy (\ref{VMspe}) with the momentum-dependent effective masses, which agrees well
with the HF result even above {the} Fermi energy as shown in Fig.~\ref{SPE1app}.
We note that the value of $k_c$ has been determined to be $2.5k_{F}$ independently of the value of $C_{dd}$ in such a way that the model single particle energy (\ref{VMspe}) is consistent with the HF calculations.

  \renewcommand{\theequation}{B.\arabic{equation} }
  \setcounter{equation}{0}
  \section{Eigenvalue problem for polaron energy} \label{Eigenvalue problem for polaron energy}
From Eq.~(\ref{eve2}), {one finds an equation for}
the auxiliary field $\chi_{\bp} = F_{0} + \sum_{k>}F_{\bk,\bp}$,
    \begin{align}
       &\chi_{\bp} = F_{0} + \frac{1}{V} \sum_{k>} \frac{g \chi_{\bp}}{E - \Omega_{\bk,\bp;\bm{P}}}.
    \end{align}
   Plugging this into  Eq.~(\ref{eve1}), one {obtains}
    \begin{align}
      & \frac{g}{V} \sum_{p} \frac{g^{-1} F_{0}}{ g^{-1} - \frac{1}{V} \sum_{k>} \frac{1}{E - \Omega_{\bk,\bp;\bm{P}}} }  = \biggr(E - \frac{ P^{2} }{2 M } \biggr) F_{0},
    \end{align}
{which} leads to Eq.~(\ref{fkp}).

  \renewcommand{\theequation}{C.\arabic{equation} }
  \setcounter{equation}{0}
  \section{Effective range} \label{effective range}
    We derive a relation between {the} momentum cutoff and {the} effective range. From the LS equation, the two-body $T$-matrix {satisfies}
    \begin{align}
      T(\bk, \bk', \omega_{+}) = g \gamma_{\bk} \gamma_{\bk'} + \frac{g}{V} \sum_{\bp} \frac{\gamma^{2}_{\bp} }{\omega_{+} - \bp^{2}/2m_{r} } T(\bp, \bk', \omega_{+}),
    \end{align}
    where we take {a} cutoff $\Lambda$ {via}
    \begin{align}
      \gamma_{\bk} = \theta(\Lambda - |\bk|).
    \end{align}

    If we set
    \begin{align}
      T(\bk, \bk', \omega_{+}) = \gamma_{\bk} t(\omega_{+}) \gamma_{\bk'}, \label{tmat}
    \end{align}
    then
    \begin{align}
      t(\omega_{+}) = g \biggr[ 1 - \frac{g}{V} \sum_{\bp}^{\Lambda} \frac{1}{\omega_{+} - p^{2}/2m_{r} } \biggr]^{-1}. \label{tomega}
    \end{align}

    {Since} the scattering amplitude $f(\bk)$ {is related to the $T$-matrix as}
    \begin{align}
      f(\bk) = -\frac{m_{r}}{2 \pi} T(\bk,\bk,k^{2}/2m_{r} + i \delta),
    \end{align}
    {we obtain} from Eqs.\ (\ref{tmat}) and (\ref{tomega})
    \begin{align}
      f(\bk) &= -\frac{m_{r}}{2 \pi} \gamma_{\bk} g \biggr[ 1 + \frac{g}{V} \sum_{\bp}^{\Lambda} \frac{1}{p^{2}/2m_{r} - k^{2}/2m_{r} - i \delta } \biggr]^{-1} \nonumber
      \\
      &= -\frac{m_{r}}{2 \pi} \gamma_{\bk}  g \biggr[1 + \frac{g m_{r}}{\pi^{2}} \biggr( \Lambda - k \tanh^{-1} \biggr( \frac{k}{\Lambda} \biggr)  +\frac{i\pi k}{2} \biggr)   \biggr]^{-1},
    \end{align}
    where we have used
    \begin{align}
      \frac{1}{V}\sum_{\bp}^{\Lambda} \frac{1}{p^{2}/2m_{r} - k^{2}/2m_{r} - i \delta } &= \frac{2m_{r}}{2\pi^{2}} \int_{0}^{\Lambda} dp \frac{p^{2}}{p^{2} -  k^{2} - i \delta} \nonumber
      \\
      &= \frac{m_{r}}{\pi^{2}} \biggr( \Lambda - k \tanh^{-1} \biggr( \frac{k}{\Lambda} \biggr) +\frac{i\pi k}{2}   \biggr).
    \end{align}

    The scattering amplitude {can be} also written {in terms of the} $s$-wave scattering length $a$ and effective range $r_{e}$ as
    \begin{align}
      f(\bk) = -\frac{1}{a^{-1} - r_{e} k^{2} / 2 + ik},
    \end{align}
    so we {obtain}
    \begin{align}
      -a^{-1} + \frac{1}{2}r_{e}k^{2} = - \theta(\Lambda - |\bk|) \frac{2 \pi}{m_{r}} \biggr[ g^{-1} + \frac{m_{r}}{\pi^{2}} \biggr( \Lambda - k \tanh^{-1} \biggr( \frac{k}{\Lambda} \biggr) \biggr) \biggr]. \label{recut}
    \end{align}

    Finally, {comparison of the $O(k^{2})$ terms in the left and right sides} of Eq.\ (\ref{recut}),
   {which can be made by using} $\tanh^{-1}(k/\Lambda) \approx k/\Lambda + O(k^{3})$,
    {leads to} the relation between {the} effective range and {the} cutoff,
    \begin{align}
      r_{e} = \frac{4}{\pi \Lambda}.
    \end{align}

\end{document}